\documentclass[1p,preprint,10pt]{elsarticle}
\usepackage{amsmath}
\usepackage{amssymb}
\usepackage{amsmath}
\usepackage{textcomp}
\usepackage{color}
\usepackage{framed}
\usepackage[usenames,dvipsnames]{xcolor}
\usepackage{hyperref}
\hypersetup{colorlinks=true,breaklinks=true,linkcolor=blue,urlcolor=blue,citecolor=blue}
\usepackage{adjustbox}
\usepackage{tikz}
\usepackage{float}
\usetikzlibrary{shapes,arrows,positioning,fit}
\usetikzlibrary{calc}

\tikzstyle{decision} = [diamond, draw, fill=green!20,
    text width=4.5em, text badly centered, node distance=3cm, inner sep=0pt]
\tikzstyle{block} = [rectangle, draw, fill=blue!20,
    text width=5em, text centered, rounded corners, minimum height=2em]
\tikzstyle{terminal} = [ellipse, draw, fill=red!20,
    text width=5em, text centered, minimum height=2em]
\tikzstyle{line} = [draw, -latex']

\makeatletter
\tikzset{
  fitting node/.style={
    inner sep=0pt,
    fill=none,
    draw=none,
    reset transform,
    fit={(\pgf@pathminx,\pgf@pathminy) (\pgf@pathmaxx,\pgf@pathmaxy)}
  },
  reset transform/.code={\pgftransformreset}
}
\makeatother

\journal{Computer Physics Communications}

\usepackage{lineno}

\definecolor{darkblue}{rgb}{0,0,.6}
\definecolor{darkred}{rgb}{.6,0,0}
\definecolor{darkgreen}{rgb}{0,.6,0}
\definecolor{red}{rgb}{.98,0,0}

\def\ssmall{\fontsize{8pt}{2pt}\selectfont}

\usepackage{listings}
\usepackage{caption}

\DeclareCaptionFormat{listing}{\rule{\dimexpr\textwidth+17pt\relax}{0.4pt}\par\vskip1pt#1#2#3}
\lstnewenvironment{cpplisting}[1][]
  {\lstset{language=C++,numbers=right,#1}}{}

\lstnewenvironment{pylisting}[1][]
  {\lstset{language=Python,numbers=right,#1}}{}

\newcommand\py[1]{\lstinline[language=Python]{#1}}
\newcommand\fo[1]{\lstinline[language=Fortran]{#1}}

\newcounter{bla}

\begin{document}

\begin{frontmatter}

\title{TRIQS/DFTTools: A TRIQS application for {\it ab initio} calculations of correlated materials}

\author[ITPCP]{Markus~Aichhorn\corref{author}} \ead{aichhorn@tugraz.at}
\author[X,CDF]{Leonid~Pourovskii} \ead{leonid@cpht.polytechnique.fr}
\author[X,CEA]{Priyanka~Seth} \ead{priyanka.seth@cea.fr}
\author[ARG]{Veronica~Vildosola} \ead{vildosol@tandar.cnea.gov.ar}
\author[ITPCP]{Manuel~Zingl} \ead{manuel.zingl@tugraz.at}
\author[GEN]{Oleg~E.~Peil} \ead{oleg.peil@unige.ch}
\author[RUT]{Xiaoyu~Deng} \ead{xiaoyu.deng@gmail.com}
\author[IJS]{Jernej~Mravlje} \ead{jernej.mravlje@ijs.si}
\author[ITPCP]{Gernot~J.~Kraberger} \ead{gkraberger@tugraz.at}
\author[TOUL]{Cyril~Martins} \ead{cyril.martins@irsamc.ups-tlse.fr}
\author[X,CDF]{Michel~Ferrero} \ead{michel.ferrero@polytechnique.edu}
\author[CEA]{Olivier~Parcollet} \ead{olivier.parcollet@cea.fr}

\cortext[author] {Corresponding author}

\address[ITPCP]{Institute of Theoretical and Computational Physics, TU
  Graz, NAWI Graz, Petersgasse 16, 8010 Graz, Austria}
\address[X]{Centre de Physique Th\'eorique, \'Ecole Polytechnique,
  CNRS, Universit\'e Paris-Saclay, F-91128 Palaiseau, France}
\address[CDF]{Coll\`ege de France, 11 place Marcelin Berthelot,
  75005 Paris, France}
\address[CEA]{Institut de Physique Th\'eorique (IPhT), CEA, CNRS, UMR CNRS 3681, 91191 Gif-sur-Yvette, France}
\address[ARG]{Departamento de F\'isica de la Materia Condensada, GIyA, CNEA (1650) San
Mart\'in, Provincia de Buenos Aires and Consejo Nacional de
Investigaciones Cient\'ificas y T\'ecnicas (CONICET), Ciudad Aut\'onoma de
Buenos Aires, Argentina}
\address[GEN]{Department of Quantum Matter Physics, University of Geneva, 24 Quai Ernest-Ansermet, 1211 Geneva 4, Switzerland}
\address[RUT]{Department of Physics and Astronomy, Rutgers University, Piscataway, New Jersey 08854, USA}
\address[IJS]{Jozef Stefan Institute, Jamova 39, Ljubljana, Slovenia}
\address[TOUL]{Laboratoire de Chimie et Physique Quantiques, UMR 5626, IRSAMC, CNRS et Universit\'e de Toulouse (UPS), 118 route de Narbonne, 31062 Toulouse, France}

\begin{abstract}

We present the \py{TRIQS/DFTTools} package, an application based on the TRIQS
library that connects this toolbox to realistic materials calculations based on
density functional theory (DFT). In particular, \py{TRIQS/DFTTools} together with
TRIQS allows an efficient implementation of DFT plus dynamical mean-field theory
(DMFT) calculations.  It supplies tools and methods to construct Wannier
functions and to perform the DMFT self-consistency cycle in this
basis set. Post-processing tools, such as band-structure plotting or the
calculation of transport properties are also implemented.
The package comes with a fully charge self-consistent interface to the Wien2k
band structure code, as well as a generic interface
that allows to
use \py{TRIQS/DFTTools} together with a large variety of DFT codes. It is distributed
under the GNU General Public License (GPLv3).

\end{abstract}

\end{frontmatter}


\noindent {\bf PROGRAM SUMMARY}

\begin{small}
\noindent
{\em Program Title:} TRIQS/DFTTools                                      \\
{\em Project homepage:} https://triqs.ipht.cnrs.fr/applications/dft\_tools               \\
{\em Catalogue identifier:} --                                  \\
{\em Journal Reference:} --                                     \\
{\em Operating system:} Unix, Linux, OSX \\
{\em Programming language:} \verb*#Fortran#/\verb*#Python#\\
{\em Computers:} 
  Any architecture with suitable compilers including PCs and clusters \\
{\em RAM:} Highly problem dependent \\
{\em Distribution format:} GitHub, downloadable as zip \\
{\em Licensing provisions:} GNU General Public License (GPLv3)\\
{\em Classification:} 6.5, 7.3, 7.7, 7.9 
{\em External routines/libraries:} 
  \verb#TRIQS#, 
  \verb#cmake#.\\ 
{\em Running time:} Tests take less than a minute; otherwise highly problem dependent.\\
{\em Nature of problem:}\\
Setting up state-of-the-art methods for an {\it ab initio} description of
correlated systems from scratch requires a lot of code development. In order to
make these calculations possible for a larger community there is need for
high-level methods that allow to construct DFT+DMFT calculations in a modular
and efficient way.  \\
{\em Solution method:}\\
We present a \verb#Fortran#/\verb#Python# open-source computational library that
provides high-level abstractions and modules for the combination of DFT with
many-body methods, in particular the dynamical mean-field theory. It allows the
user to perform fully-fledged DFT+DMFT calculations using simple and short
Python scripts. 

\end{small}

\section{Introduction and Motivation}
\label{sec:introduction}

When describing the physical and also chemical properties of crystalline
materials, there is a standard method that is used with great success for a
large variety of systems: density functional theory (DFT). This powerful theory
states that the electron density of a system uniquely determines all
ground-state properties. 
However, in
order to find the electron density in practice one usually resorts to 
the Kohn-Sham scheme involving an approximate exchange-correlation potential, such as
the local-density or the generalized-gradient
approximations. The solution of the self-consistent Kohn-Sham
equations yields
the Kohn-Sham orbitals (also called Bloch bands)
$|\psi_{\mathbf{k}\nu}\rangle$ and the corresponding Kohn-Sham energies
$\varepsilon_{\mathbf{k}\nu}$. The index $\nu$ labels the electronic band, and
$\mathbf{k}$ is the wave vector within the first Brillouin zone
(BZ). Below, we will always imply the Kohn-Sham scheme when talking
about DFT.

The solutions of the Kohn-Sham equations form bands of allowed states in
momentum space. These states are filled up to the Fermi level by the electrons
according to the Pauli principle, and one can use this simple picture for
example to explain the electronic band structure of materials such as
elementary copper or aluminium.  Following this principle one can try to
classify all existing materials into metals and insulators. A system is a metal
if there are an odd number of electrons per unit cell in the valence bands, since
this leads to a partially filled band that cuts the Fermi energy and thus
produces a Fermi surface. On the other hand, an even number of electrons per
unit cell can lead to completely filled bands with a finite excitation gap to the
conduction bands, i.e. insulating behaviour. 

However, there are certain compounds where this classification into metals and
insulators fails dramatically. This happens in particular in systems with
partially filled $d$- and $f$-shells. There, DFT predicts metallic behaviour due
to the open-shell setting, but in experiments many of these materials actually
show insulating behaviour. This cannot be explained by band theory and the Pauli
principle alone, and a different mechanism has to be invoked. The insulating
behaviour in such systems, among which are the Mott insulators~\cite{RevModPhys.70.1039},
comes by solely due to correlations between the electrons. 
Moreover, many interesting
properties, such as transport, photoemission, etc. are temperature-dependent,
which cannot be treated with a ground-state theory such as DFT alone.

In order to describe correlated materials within \textit{ab initio} schemes one has
to go beyond a standard DFT treatment. One step in this direction is the so
called DFT+U approximation~\cite{ldau} which can induce the opening of a gap through a
static shift of the Kohn-Sham eigenvalues. However, this technique cannot
properly describe the physics of correlated metals, whose low-energy properties
are dominated by dynamical many-body effects.  In the mid-nineties, a very
successful method was introduced to tackle this problem, namely the combination
of DFT with the dynamical mean-field theory
(DMFT)~\cite{DMFTRMP,Anisimov1997,RevModPhys.78.865}. The main idea is to add local quantum
correlations in the framework of DMFT to the non-local quantum Hamiltonian
obtained by DFT. This method has been since applied to a large variety of
problems, including high-temperature superconductors, organic insulators, iron
pnictides, oxide perovskites, and other strongly-correlated materials.

Since the first proposal of DFT+DMFT, there has been a tremendous amount of new
developments of numerical tools, including (fully charge self-consistent)
interfaces between DFT codes and the
DMFT~\cite{pourovskii1,PhysRevB.71.125119,lechermann_wannier-bis,haule2009,w2w,amadon2012,PhysRevB.90.235103},
and exact algorithms for solving the DMFT
equations~\cite{rubtsov_continuous-time_2005,PhysRevLett.97.076405,PhysRevB.74.155107,PhysRevB.75.155113,RevModPhys.83.349,cthyb_2015}. 
As a result, many more classes of materials can be studied within this framework
on an \textit{ab initio} basis, with much higher accuracy than just a decade
ago. Importantly, these technical advances allow studies of realistic
materials at experimentally interesting ambient and even lower
temperatures (see, e.g., Ref.~\cite{mravlje11} for Sr$_2$RuO$_4$), 
in contrast to the older techniques that allowed studies down to
temperatures of 1000\,K only (see, e.g., Ref.~\cite{pchelkina} for the
same material as above). 
The drawback of these new methods, however, is that they require a
substantial amount of coding effort to keep up with the state-of-the-art
techniques in the field.

In this paper we present version 1.3 of the \py{TRIQS/DFTTools}
application, which is based on the TRIQS
library~\cite{triqs_2015}. The purpose of this application is to
provide a complete set of tools to perform 
fully charge self-consistent \textit{ab initio} calculations for correlated
materials, and it enables researchers in the field of correlated materials
to do these calculations 
without dedicating time to code their
own implementation. The \py{TRIQS/DFTTools} package provides a fully charge
self-consistent interface to the Wien2k~\cite{wien2k} band structure
program, as well as an interface that is usable with many other DFT or tight-binding
codes, however, without the possibility of full charge self-consistency.
Physical properties that can be calculated range from spectral functions and
(projected) densities of states (DOS), charges and magnetic moments, to more
involved quantities like optical conductivity and thermopower.  The package can
handle spin-polarised cases, as well as spin-orbit coupled
systems~\cite{martins2011}.  It is released under the Free Software GPLv3
license.

The paper is organised as follows. We start with a brief general discussion of
the structure of \py{TRIQS/DFTTools} in Sec.~\ref{sec:structure}. In
Sec.~\ref{sec:interface} we show how the interface between DFT and DMFT is set
up, in particular for the Wien2k DFT package. We will also introduce the
projective Wannier functions that are used as the basis set for the DMFT
calculations. We continue in Sec.~\ref{sec:one-shot} with a simple example of a
DFT+DMFT calculation, without charge self-consistency. In Sec.~\ref{sec:fullsc}
the concept of full charge self-consistency is introduced, followed by a
discussion of post-processing tools in Sec.~\ref{sec:postprocessing}.
Information on how to obtain and install the code is given in
Sec.~\ref{sec:starting}, and we finally conclude in Sec.~\ref{sec:summary}.

\section{General structure of \py{TRIQS/DFTTools}}
\label{sec:structure}

\begin{figure}[t]
  \centering
  \includegraphics[width=0.8\textwidth]{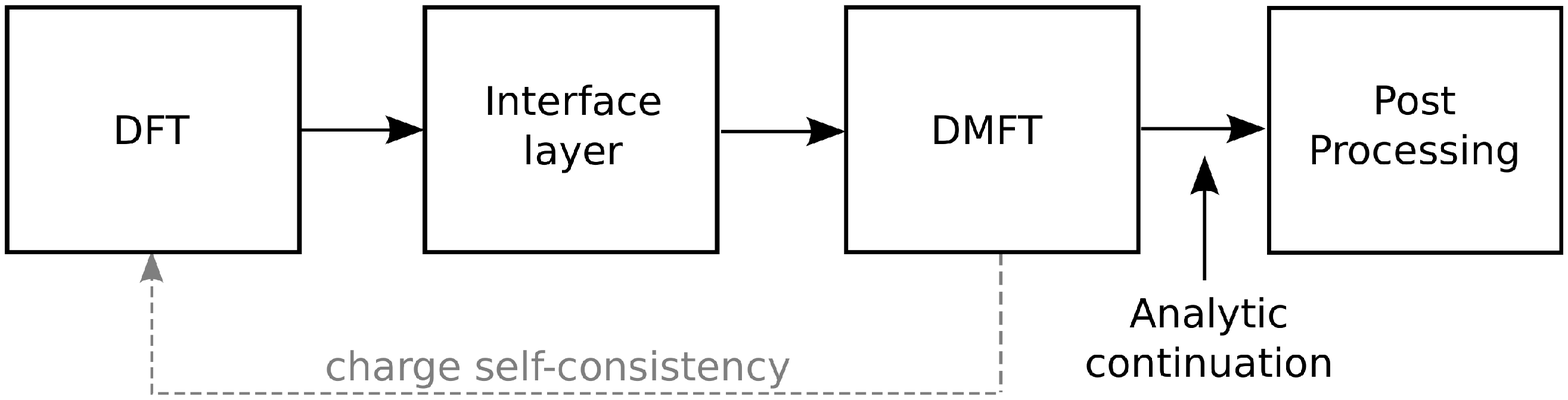}
  \caption{ \label{fig:structure}
    General structure of the \py{TRIQS/DFTTools} package.}
\end{figure}

The central part of \py{TRIQS/DFTTools} is a collection of Python modules that allow
an efficient implementation of the DMFT self-consistency cycle. Since the input
for this DMFT part has to be provided by DFT calculations, there needs to be
another layer that connects the Python-based modules with the DFT output.
Naturally, this interface depends on the DFT package at hand. Within
\py{TRIQS/DFTTools} we provide an interface to the Wien2k band structure package, and
a general interface that can be used in a more general setup. Note that only the
Wien2k interface allows fully charge self-consistent calculations. Support for
other codes is in progress. 

The basic structure is schematically presented in Fig.~\ref{fig:structure}.
After performing a DFT calculation, the Kohn-Sham orbitals are used to construct
localised Wannier orbitals, and all required information is converted into an
hdf5 file. This file is used by the Python modules of
\py{TRIQS/DFTTools} to perform the DMFT calculation. The fully charge
self-consistent loop can be 
closed by taking the interacting density matrix and using it to recalculate the
ground state density of the crystal in Wien2k. This leads to a new Kohn-Sham
exchange-correlation potential, and thus to new orbitals. The full loop has to be
repeated until convergence.

\py{TRIQS/DFTTools} also provides tools for analysing the data, such as
spectral functions, densities of states, and transport properties. Depending on the
method of solution of the Anderson impurity problem within the DMFT loop, the
user may need to perform analytic continuation of the self-energy in order to
use these tools\footnote{Note that \py{TRIQS/DFTTools} itself does not provide any
method to perform the analytic continuation.}. 

\section{The interface to DFT}
\label{sec:interface}

\subsection{Projective Wannier functions}

In order to perform the DMFT calculations one has to choose an appropriate localised
basis set. In this package, we provide a method to calculate projective Wannier
functions $|w_{\mathbf{k}m}^{\alpha,\sigma}\rangle$ based on the Wien2k
Kohn-Sham orbitals, as introduced in Ref.~\cite{triqs_wien2k_interface}. We
denote orbitals in the correlated subspace $\mathcal{C}$ where many-body
correlations are applied by Latin letters $m$, the index of the atom in the unit
cell by $\alpha$, and the spin projection by $\sigma$.

The basic idea of the projective Wannier function technique
\cite{anisimov_projective} is to expand a set of local orbitals
$|\chi_m^{\alpha,\sigma}\rangle$ over a restricted number of Bloch bands
\begin{equation}
|\tilde\chi_{\mathbf{k}m}^{\alpha,\sigma}\rangle
=\sum_{\nu\in \mathcal{W}}\langle\psi_{\mathbf{k}\nu}^{\sigma}|\chi_{m}^{\alpha,\sigma}\rangle
|\psi_{\mathbf{k}\nu}^{\sigma}\rangle,\label{eq:Correl-orb-1}
\end{equation}
where $\nu$ is the band index and the sum is carried out over Bloch bands within
an energy window $\mathcal{W}$. Note that the functions
$|\tilde\chi_{\mathbf{k}m}^{\alpha,\sigma}\rangle$ are not orthonormal due to
this truncation.

The orbitals $\left|\tilde\chi_{\mathbf{k}m}^{\alpha,\sigma} \right\rangle $ can
be orthonormalised, giving a set of Wannier functions:
\begin{equation}
  \left|w_{\mathbf{k}m}^{\alpha,\sigma}\right\rangle
  =\sum_{\alpha',m'}S_{m,m'}^{\alpha,\alpha'}
  |\tilde \chi_{\mathbf{k}m'}^{\alpha',\sigma}\rangle ,
  \label{eq:wannier}
\end{equation}
where $S_{m,m'}^{\alpha,\alpha'} =\left\{ O(\mathbf{k},\sigma)^{-1/2}\right\}
_{m,m'}^{\alpha,\alpha'}$ and
$O_{m,m'}^{\alpha,\alpha'}(\mathbf{k},\sigma)=\langle \tilde
\chi_{\mathbf{k}m}^{\alpha,\sigma}|\tilde\chi_{\mathbf{k}m'}^{\alpha',\sigma}
\rangle $ are the overlap matrix elements. The functions
$\left|w_{\mathbf{k}m}^{\alpha,\sigma}\right\rangle$ are, in fact, the Bloch
transform of the corresponding real-space Wannier functions
$\left|w_{\mathbf{R}m}^{\alpha,\sigma}\right\rangle$, through
$\left|w_{\mathbf{k}m}^{\alpha,\sigma}\right\rangle=\sum_{\mathbf{R}}
e^{i\mathbf{kR}}\left|w_{\mathbf{R}m}^{\alpha,\sigma}\right\rangle$. 

\begin{figure}[t]
   \centering
   \includegraphics[width=0.6\textwidth]{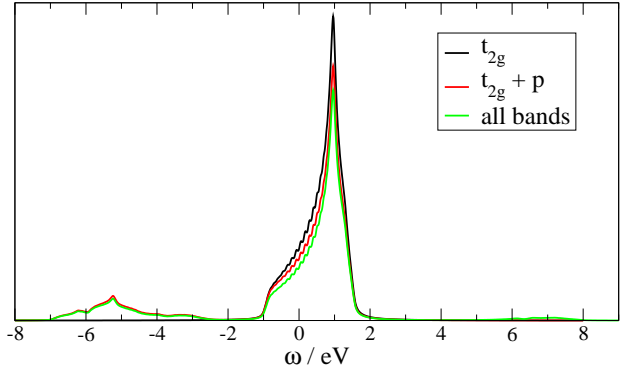}
   \caption{\label{fig:wanniers}
     DOS of the vanadium t$_{2g}$-like Wannier functions of SrVO$_3$ produced by
     \fo{dmftproj}, for 3 different choices of projection window. 
     Black line: only vanadium t$_{2g}$ bands, $\mathcal{W}=[-2.0,2.0]$\,eV. 
     Red line: vanadium t$_{2g}$ and oxygen $p$ bands, $\mathcal{W}=[-8.0,2.0]$\,eV. 
     Green line: even larger window, $\mathcal{W}=[-8.0,8.0]$\,eV. 
     Note in the latter case the small additional weight around 7\,eV.}
\end{figure}

\begin{figure}[t]
   \centering
   \includegraphics[width=0.8\textwidth]{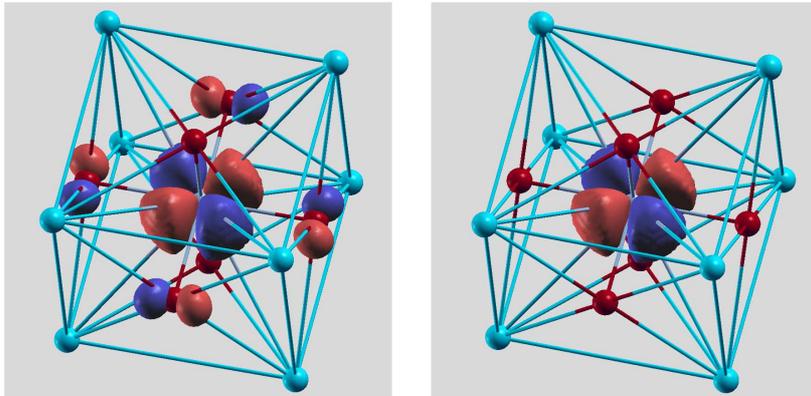}
   \caption{\label{fig:wanniers_plot}
	 Real-space representation of the vanadium t$_{2g}$ Wannier function of
     SrVO$_3$. Left plot: projection only to t$_{2g}$ bands. Right plot: projection
     to all vanadium $d$ and oxygen $p$ bands.  The blue and red colours indicate the
     negative and positive phases, respectively, of the Wannier function. Note the
     much better localisation of the $d$-like Wannier functions in the latter case,
     without weight around the oxygen positions.  The Wannier functions are
     constructed using the \fo{dmftproj} program, their real-space representations
     are generated with the help of the \fo{wplot} program from the
     wien2wannier~\cite{w2w} package. Plots are produced using
     XCrysDen~\cite{xcrysden}.}   
\end{figure}

In practice it is much more convenient to work with projection operators, which
transform quantities from the Bloch band basis to the Wannier orbital basis.
We define these operators as
\begin{equation}
  \hat P^{\alpha,\sigma}(\mathbf k) = \sum_{m\in \mathcal
    C}|w_{\mathbf{k}m}^{\alpha,\sigma}\rangle\langle
  w_{\mathbf{k}m}^{\alpha,\sigma}|.\label{eq:projop}
\end{equation}
The matrix elements of these operators are calculated straight-forwardly using
the above Wannier functions: 
\begin{align}
  P_{m\nu}^{\alpha,\sigma}
  (\mathbf{k})&=\underset{\alpha'm'}{\sum}S_{m,m'}^{\alpha,\alpha'}
  \widetilde{P}_{m'\nu}^{\alpha',\sigma}(\mathbf{k})\label{eq:wannier-proj}\\
  \widetilde P_{m\nu}^{\alpha,\sigma}(\mathbf{k})&=\langle
    \tilde\chi_{m}^{\alpha,\sigma}|\psi_{\mathbf{k}\nu}^{\sigma}\rangle,
    \qquad \nu\in\mathcal{W}
    \label{eq:proj-1}
\end{align}
These projectors can now be used to calculate, e.g., the local projected 
non-interacting Matsubara Green's function from the DFT Green's function,
\begin{equation}
G_{mn}^{0,\alpha}(i\omega_n) = \sum_\mathbf{k}\sum_\nu P_{m\nu}^{\alpha}
  (\mathbf{k}) \frac{1}{i\omega_n -\varepsilon_{\mathbf{k}\nu}+\mu}
  P_{n\nu}^{\alpha*} (\mathbf{k}),
\end{equation}
where we have dropped the spin index for better readability.  
Note that throughout the paper we assume a proper normalisation of the momentum sum over 
$\mathbf{k}$ in the first Brillouin zone, i.e. $\sum_{\mathbf{k}}1=1$.
From this
non-interacting Green's function one can calculate the density of states. As an
example, we show in Fig.~\ref{fig:wanniers} the DOS for the prototypical
material SrVO$_3$. Three different Wannier constructions are displayed. First, a
projection where only the vanadium t$_{2g}$ bands are taken into account is
shown in black.  The projection that comprises vanadium t$_{2g}$ as well as the
oxygen $p$ bands is shown in red, and the green line is the DOS for a projection
using DFT bands up to 8\,eV. The difference in the latter two projections is
minor, and consists primarily in the small transfer of weight to large energies
around 7\,eV.

In Fig.~\ref{fig:wanniers_plot} we show the real-space representation of the
Wannier charge density. The left plot shows a t$_{2g}$-like Wannier function for
a projection of t$_{2g}$ bands only, which corresponds to the black line in
Fig.~\ref{fig:wanniers}. The right plot is the corresponding Wannier function
when all bands are included in the projection. The effect of increasing the
projection window is obvious: it results in a better localisation of the
resulting Wannier functions. 

All necessary steps to construct the orthonormalised set of projection operators
from Wien2k DFT calculations is contained in the program \fo{dmftproj} that is
shipped with \py{TRIQS/DFTTools}.  In Wien2k the Bloch functions are expanded in the
LAPW basis set as
\begin{equation}
  \begin{split}
    \psi_{\mathbf{k}\nu}^{\sigma}(\mathbf{r})&=\overset{N_{PW}}{\underset{\mathbf{G}}{\sum}}
    c_{\mathbf{G}}^{\nu,\sigma}(\mathbf{k})\underset{lm}{\sum}A_{lm}^{\alpha,\mathbf{k}+
      \mathbf{G}}u_{l}^{\alpha,\sigma}(r,E_{1l}^{\alpha})Y_{m}^{l}(\hat{r})\\
    &+\overset{N_{lo}}{\underset{n_{lo}=1}{\sum}}c_{lo}^{\nu,\sigma}
    \left[A_{lm}^{\alpha,
        lo}u_{l}^{\alpha,\sigma}(r,E_{1l}^{\alpha})+B_{lm}^{\alpha,lo}
      \dot{u}_{l}^{\alpha,\sigma}(r,E_{1l}^{\alpha})\right]Y_{m}^{l}(\hat{r})+\\
    &+\overset{N_{LO}}{\underset{n_{LO}=1}{\sum}}c_{LO}^{\nu,\sigma}
    \left[A_{lm}^{\alpha,
        LO}u_{l}^{\alpha,\sigma}(r,E_{1l}^{\alpha})+C_{lm}^{\alpha,LO}u_{l}^{\alpha,\sigma}(r,E_{2l}^{\alpha})\right]Y_{m}^{l}(\hat{r}),
  \end{split}
  \label{eq:KS-eigen-2}
\end{equation}
where $N_{PW}$ is the total number of plane waves in the basis set, $N_{lo}$ is
the number of local orbitals, and $N_{LO}$ the corresponding number of auxiliary
orbitals for semi-core states.  For more details on the Wien2k basis set we
refer the reader to the
literature~\cite{wien2k,LAPWSingh,triqs_wien2k_interface}.

As initial localised orbitals $|\chi_m^{\alpha,\sigma}\rangle$ in our projective
Wannier function construction, we choose the solutions of the Schr\"{o}dinger
equation within the muffin-tin sphere
$\left|u_{l}^{\alpha,\sigma}(E_{1l})Y_{m}^{l}\right\rangle$ at the corresponding
linearisation energy $E_{1l}$. Inserting this ansatz into
Eq.~\ref{eq:Correl-orb-1} and using the orthogonality relations of the solutions
of the Schr\"odinger equation allows to calculate the matrix elements of the
projection operators. After some algebra one arrives at the expressions

\begin{equation}
  \begin{split}
    \widetilde P_{m\nu}^{\alpha,\sigma}(\mathbf{k})&=\left\langle u_{l}^{\alpha,\sigma}
      (E_{1l})Y_{m}^{l}\right.\left|\psi_{\mathbf{k}\nu}^{\sigma}
    \right\rangle\\
    &=A_{lm}^{\nu,\alpha}(\mathbf{k},\sigma)
    +\overset{}{\underset{n_{LO}=1}{\sum}}
    C_{lm,LO}^{\nu,\alpha}(\mathbf{k},\sigma),
  \end{split}
  \label{eq:proj-2}
\end{equation}
with contributions from the LAPW and/or APW+$lo$ orbitals
\begin{equation}
  \begin{array}{ll}
    A_{lm}^{\nu,\alpha}(\mathbf{k},\sigma)&=\overset{N_{PW}}
    {\underset{\mathbf{G}}{\sum}}c_{\mathbf{G}}^{\nu,\sigma}(\mathbf{k})
    A_{lm}^{\alpha,\mathbf{k}+\mathbf{G}}\\
    &+\overset{N_{lo}}{\underset{n_{lo}=1}{\sum}}c_{lo}^{\nu,\sigma}
    A_{lm}^{\alpha,lo}+\overset{N_{LO}}{\underset{n_{LO}=1}{\sum}}c_{LO}^{\nu,\sigma}
    A_{lm}^{\alpha,LO}
  \end{array}
  \label{eq:coeff-proj-1}
\end{equation}
as well as the contribution from the semi-core orbitals,
\begin{equation}
  C_{lm,LO}^{\nu,\alpha}(\mathbf{k},\sigma)=
  c_{LO}^{\nu,\sigma}C_{lm}^{\alpha,LO}
  \left\langle u_{l}^{\alpha,\sigma}(E_{1l})Y_{m}^{l}|u_{l'}^{\alpha,\sigma}(E_{l,LO})
    Y_{m'}^{l'}\right\rangle.
  \label{eq:coeff-proj-2}
\end{equation}

\fo{dmftproj} takes all the necessary data from Wien2k, evaluates
Eqs.~\ref{eq:proj-2} and~\ref{eq:coeff-proj-2} to obtain the matrix elements of
the projection operators, Eqs.~\ref{eq:wannier-proj} and~\ref{eq:proj-1}.
Further information needed for the DMFT calculation are i) the rotation matrix
from local to global coordinate systems, ii) the Kohn-Sham Hamiltonian within
the projection window, and iii) the symmetry operation matrices for the BZ
integration. All this information is extracted by \fo{dmftproj} from the Wien2k
output.

\subsection{Usage}

As a prerequisite a self-consistent Wien2k DFT calculation has to be performed.
At this point, the further steps for the interface are
\begin{itemize}
\item writing of necessary data from the Wien2k calculation into
  files, which is done using the Wien2k command \textit{x lapw2 -almd},
\item running \fo{dmftproj} to calculate the projective Wannier functions,
\item conversion of the text output files into the hdf5 file format using the Python
      modules provided.
\end{itemize}

After these steps, all necessary information for a DMFT calculation is stored in
a single hdf5 file. For the details of these steps, and all input/output
parameters we refer the reader to the extensive online documentation of
\py{TRIQS/DFTTools} and the example below.

\subsection{General interface for non-Wien2k users}

The \py{TRIQS/DFTTools} package also contains an interface layer that can be used in
many situations, albeit without the possibility of full charge self-consistency.
In fact, as an input it takes the correlated subspace Hamiltonian
$H_{mn}(\mathbf{k})$, where $m$ and $n$ are orbital indices, and the
$\mathbf{k}$-mesh spans the first BZ. Unlike the Wien2k converter, the
calculation is done exclusively in Wannier orbital space.  The converter takes
this $H_{mn}(\mathbf{k})$, together with information such as the quantum numbers
of the correlated orbitals and the required electron density, and produces a
hdf5 file that can be used for the DMFT calculations. We note that this
interface is completely independent of how the $H_{mn}(\mathbf{k})$ is produced.
It can use, for instance, output from
Wannier90~\cite{PhysRevB.56.12847,PhysRevB.65.035109}, which is available for
many DFT codes, from NMTO calculations~\cite{PhysRevB.62.R16219}, or from any
other tight-binding model representing the non-interacting electronic structure
of the system.

\section{One-shot DFT+DMFT calculations}
\label{sec:one-shot}

\begin{lstlisting}[language=Python,
		 caption={Sketch of the Python script for the DMFT self-consistency
                   loop in one-shot DFT+DMFT calculations. Note that
                   the solver imports, construction and execution must
                   be appropriately added and modified.} , 
                 label=oneshot,numbers=left]
# imports and initialisations
from pytriqs.applications.dft.sumk_dft import SumkDFT

SK = SumkDFT(hdf_file = 'YourDFTDMFTcalculation.h5')
S = Solver(...)         # appropriate constructor parameters to be defined
n_loops = 15            # define the number of DMFT loops

for iteration_number in range(n_loops) : # start the DMFT loop

    # set the self-energy as an attribute of the class:
    SK.set_Sigma(Sigma_imp = [ S.Sigma_iw ])     
    # calculate the chemical potential:
    chemical_potential = SK.calc_mu()                 
    # extract the local Green's function:
    S.G_iw << SK.extract_G_loc()[0]                    
    # finally get G0 (input for the Solver):
    S.G0_iw << inverse(S.Sigma_iw + inverse(S.G_iw))   

    # now solve the impurity problem:
    S.solve(...)        # appropriate solve parameters to be defined

    # calculate the new double counting correction:
    dm = S.G_iw.density()                                              
    SK.calc_dc(dm, U_interact=U, J_hund=J, orb=0, use_dc_formula=0)

# save the final results into an archive
from pytriqs.archive import HDFArchive
import pytriqs.utility.mpi as mpi
if mpi.is_master_node():
    with HDFArchive('YourDFTDMFTcalculation.h5','a') as ar:
        ar['G'] = S.G_iw
        ar['Sigma'] = S.Sigma_iw
\end{lstlisting}

As mentioned in the introduction, \py{TRIQS/DFTTools} provides simple modules that
allow the compact and efficient implementation of the DMFT self-consistency
cycle.  We will illustrate how to use the package by giving a one-to-one
correspondence of the code statements to the necessary equations. In this
section we give a simple Python script for so-called `one-shot' DFT+DMFT
calculations without full charge self-consistency. Note that the construction of
the Wannier functions and the conversion procedures discussed in the previous
section have already been performed. We assume that the hdf5 archive for the
calculation, where everything is stored, is called
\textit{YourDFTDMFTcalculation.h5}.  Furthermore, we assume that a suitable
method to solve the quantum impurity problem is available, such as the
\py{TRIQS/CTHYB} solver~\cite{cthyb_2015}. Note that
\py{TRIQS/DFTTools} itself does not provide any impurity solver.

The DMFT calculation as given in Lst.~\ref{oneshot} consists of the following
steps. First, we initialise our calculation as done in lines 2 to
6 of Lst.~\ref{oneshot}. The \py{SumkDFT} module is loaded and initialised
with the previously generated hdf5 archive \textit{YourDFTDMFTcalculation.h5}.
We initialise a solver module, which is used for the solution of the Anderson
impurity model, for instance, the \py{TRIQS/CTHYB}
package~\cite{cthyb_2015}. It is important that the user is familiar
with the 
details of this solver initialisation, including set-up of the interaction
matrix, definition of the interaction parameters $U$ and $J$, etc.  We request
$n_{\rm loops}=15$ self-consistency cycles to be done. This number has of course
to be adapted to the problem at hand in order to reach convergence.  After this
initialisation step, we begin the self-consistency loop:

\begin{enumerate}

\item[1)] Impurity self-energy (line 11): 
  The command \py{SK.set_Sigma(Sigma_imp = [ S.Sigma_iw ])} takes an impurity
  self-energy $\Sigma_{mm'}^{\rm{imp}}(i\omega)$ in orbital space, and sets it as
  a member of the \py{SK} class. We assume that the object \py{S.Sigma_iw} is
  defined in the solver. At initialisation, the self-energy can be set to zero, or
  to its Hartree value to speed up convergence.

\item[2)] Chemical potential determination (line 13):
  The chemical potential is set according to the definition
  \begin{equation}
    N_{e} =
    \mathrm{Tr}\sum_\mathbf{k}\Big[(i\omega_n-\varepsilon_{\mathbf{k}\nu}+\mu)\delta_{\nu\nu'}
      - \Sigma_{\nu\nu'}\Big]^{-1}e^{i\omega_n 0^+},
  \end{equation}
  where $N_{e}$ is the required electronic density in the projection
  window. Please note that the operation $[\cdots]^{-1}$ denotes matrix
  inversions throughout this paper. The self-energy in band space is defined by 
  \begin{equation}\label{sigma}
    \Sigma_{\nu\nu'}(\mathbf{k},i\omega_{n})
    =\sum_{\alpha,mm'}P_{m\nu}^{\alpha*}
    (\mathbf{k})\left(\Sigma_{mm'}^{\rm{imp}}(i\omega_{n})-\Sigma^{\rm{DC}}\right)
    P_{m'\nu'}^{\alpha}(\mathbf{k}).
  \end{equation} 
  We use the projectors as introduced in Sec.~\ref{sec:interface}, as
  well as a double-counting correction $\Sigma^{\rm{DC}}$~\cite{ldau}. 

\item[3)] Local Green's function (line 15): The local Green's function is
  calculated as
  \begin{equation}
    G_{mm'}^{\mathrm{loc,\alpha}}(i\omega_n) =
    \sum_\mathbf{k}\sum_{\nu\nu'}P_{m\nu}^{\alpha}(\mathbf{k})\Big[(i\omega_n-\varepsilon_{\mathbf{k}\nu}+\mu)\delta_{\nu\nu'}
      - \Sigma_{\nu\nu'}(\mathbf{k},i\omega_n)\Big]^{-1}P_{m'\nu'}^{\alpha*}(\mathbf{k}),
  \end{equation}
  where $\alpha$ is the index of the correlated atom in the unit
  cell. In Lst.~\ref{oneshot} we have only one inequivalent 
  correlated atomic shell, so we need to take only $\alpha=0$.

\item[4)] New hybridisation function (line 17): We now calculate the new Weiss field for
  the impurity problem using the Dyson equation
  \begin{equation}
    \mathcal{G}_{mm'}^0(i\omega_n) = \left[\Sigma_{mm'}^{\rm
        imp}(i\omega_n) +
      \left[G_{mm'}^{\mathrm{loc}}(i\omega_n)\right]^{-1}\right]^{-1}.
  \end{equation}

\item[5)] Solve the impurity problem (line 20): We call the solver
  method \py{S.solve} with the relevant parameters. 
  Again, the user must be familiar with the usage of the solver of choice and
  its parameters, in order to produce meaningful and physically sound results. The
  solver produces the Green's function and the self-energy of the interacting
  impurity problem. Following the \py{TRIQS/CTHYB} convention, we assume for the
  example given here that the Green's function and the self-energy are stored in
  \py{S.G_iw} and \py{S.Sigma_iw}, respectively.

\item[6)] Double counting (lines 23-24): We calculate a new double
  counting (DC) correction from the impurity density matrix (line 23). As
  additional parameters one needs to provide a Hubbard interaction value $U$ and
  Hund's exchange $J$, which have to be in accordance with the definition of the
  interaction matrix. \py{TRIQS/DFTTools} offers three different DC corrections: The
  flag \py{use_dc_formula=0} corresponds to the fully-localised limit (FLL)
  correction, \py{use_dc_formula=1} is a variant of FLL that is used for
  Kanamori-type Hamiltonians~\cite{held}, and \py{use_dc_formula=2} is the
  around-mean-field (AMF) correction. 
  This step completes the self-consistency loop.

\end{enumerate}

Steps 1 to 6 are now iterated until convergence. After that, we can save the
final solution into the hdf5 archive, as done in lines 27 to 32. We want to note
here that it is sometimes useful to save all data produced by the user in a
separate subgroup, e.g. \textit{dmft\_results}, in order to have a clean
structure in the hdf5 archive. For details how to handle the hdf5 archive we
refer the reader to Ref.~\cite{triqs_2015} and the online TRIQS documentation.

We want to stress here that this simple script of about 30 lines\footnote{The
\py{Solver} initialisation will take another 10-15 lines, see
Ref.~\cite{cthyb_2015}.} of Python code is sufficient for a full DFT+DMFT
calculation, which is possible only thanks to the modular structure of
\py{TRIQS/DFTTools} and TRIQS.

\section{Full charge self-consistency}
\label{sec:fullsc}

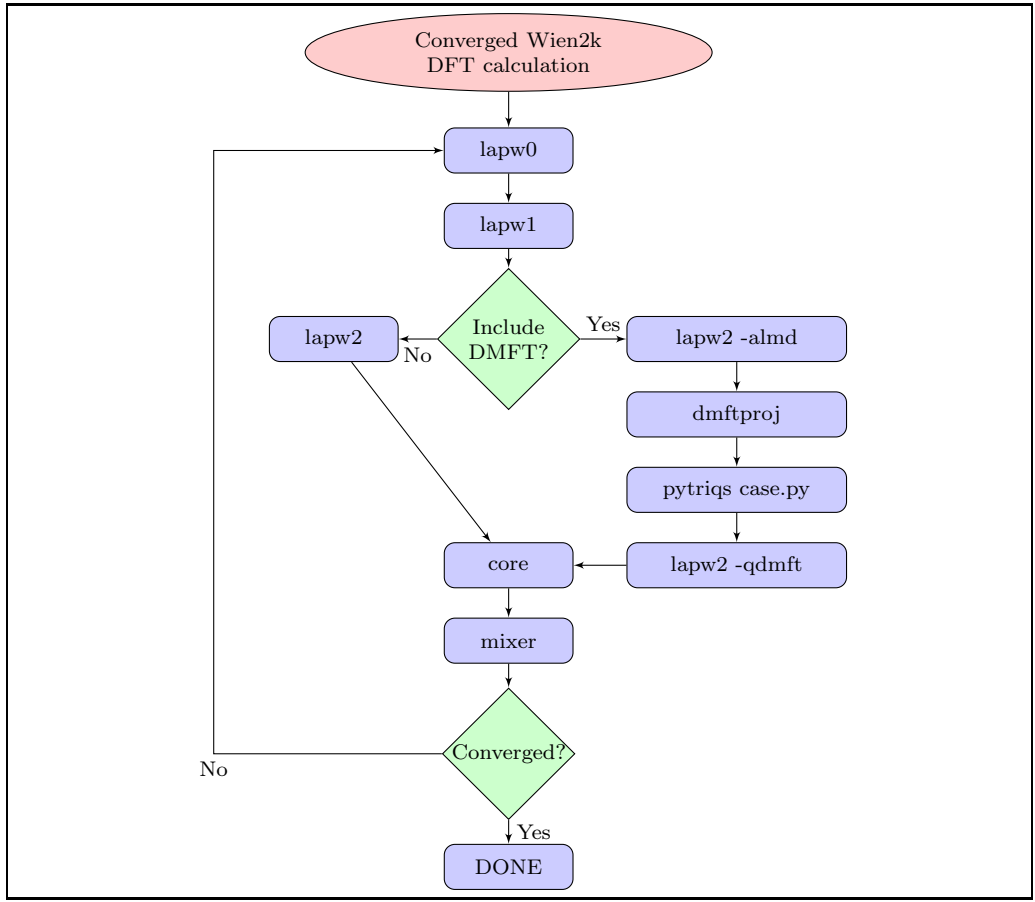
\begin{figure}[t]
    \framebox[\textwidth]{
\begin{tikzpicture}[auto]
font issue=\footnotesize
    \node [terminal, text width=12em] (input) {Converged Wien2k DFT
      calculation};
    \node [block, below of=input, node distance=1.3cm] (lapw0) {lapw0};
    \node [block, below of=lapw0, node distance=1.0cm] (lapw1) {lapw1};
    \node [decision, below of=lapw1, node distance=1.5cm] (DMFT) {Include DMFT?};
    \node [block, left of=DMFT, node distance=2.3cm] (lapw2) {lapw2};
    \node [block, right of=DMFT, text width = 9em, node distance=3.0cm] (lapw2almd) {lapw2 -almd};
    \node [block, below of=lapw2almd, text width = 9em, node distance=1.0cm] (dmftproj) {dmftproj};
    \node [block, below of=dmftproj, text width = 9em, node distance=1.0cm] (triqs) {pytriqs case.py};
    \node [block, below of=triqs, text width = 9em, node distance=1.0cm] (lapw2qdmft) {lapw2 -qdmft};
    \node [block, below of=DMFT, node distance=3cm] (core) {core};
    \node [block, below of=core, node distance=1cm] (mixer) {mixer};
    \node [decision, below of=mixer, text width=5em, node distance=1.5cm] (converged)
    {Converged?};
    \node [block, below of=converged, node distance=1.5cm] (done) {DONE};
    \path [line] (input) -- (lapw0);
    \path [line] (lapw0) -- (lapw1);
    \path [line] (lapw1) -- (DMFT);
    \path [line] (DMFT) -- node {Yes} (lapw2almd);
    \path [line] (DMFT) -- node {No} (lapw2);
    \path [line] (lapw2) -- (core);
    \path [line] (lapw2almd) -- (dmftproj);
    \path [line] (dmftproj) -- (triqs);
    \path [line] (triqs) -- (lapw2qdmft);
    \path [line] (lapw2qdmft) -- (core);
    \path [line] (core) -- (mixer);
    \path [line] (mixer) -- (converged);
    \path [line] (converged) -- node {Yes} (done);
    \path [line] (converged.west) -| node [below] {No} +(-3.0,1.5) |- (lapw0.west);
\end{tikzpicture}
}
    \caption{(Colour online) 
             Sketch of the self-consistency cycle of fully charge
             self-consistent calculations. 
             \label{fig:sccycle}}
\end{figure}

The one-shot DFT+DMFT calculation can be extended to a fully charge
self-consistent calculation with only marginal additional effort. The basic
concept is general, but the implementation will depend of course on the DFT code
that you wish to use for the calculations. We are presenting the interface to
the Wien2k DFT package, which is the first interface implemented in
\py{TRIQS/DFTTools}. We describe this interface below. Work is in progress to include
more DFT packages, in particular VASP~\cite{vasp1,vasp2}.

First of all, we want to stress that the fully charge self-consistent calculation
is controlled by Wien2k scripts. That is why a sound knowledge of Wien2k and how
it is used is absolutely necessary for the considerations in this section.

\begin{lstlisting}[language=Python,
		 caption=Code snippet for the calculation of the
                 corrected charge density due to correlations,
                 label=densitycorrection,numbers=left]
SK.set_Sigma(Sigma_imp = [ S.Sigma_iw ])
chemical_potential = SK.calc_mu( precision = 0.000001 )
dN, d = SK.calc_density_correction(filename = 'case.qdmft')
SK.save(['chemical_potential','dc_imp','dc_energ'])
\end{lstlisting}

As shown in Fig.~\ref{fig:sccycle}, the idea is to extend the existing Wien2k
self-consistency loop by including a DMFT step. Instead of simply calculating
the charge density from the Kohn-Sham orbitals as in standard DFT
(\textit{lapw2} step in Wien2k), we supplement this charge density with
correlation effects. Therefore, the Python scripts that control the DMFT
iteration have to be modified:
\begin{itemize}
\item We do not start each DMFT iteration from scratch, but instead from
  the previous iteration. This is most conveniently done by storing
  all necessary data in the hdf5 archive.
\item One DMFT iteration ($n_{\rm loops} = 1$) per full
  self-consistency cycle is usually sufficient.
  In cases where convergence is problematic, more iterations may be needed.
\item In order to calculate the correction to the charge density, one has to
  calculate the density matrix from the correlated Green's function at the end of
  the DMFT step, as shown in Lst.~\ref{densitycorrection}. The first two lines
  set the chemical potential more accurately. The important step is line 3, where
  the correlated density matrix is calculated in Bloch band space
  \begin{align}
     N^{k}_{\nu\nu'} &= \sum_{n}
     G_{\nu\nu'}(\mathbf{k},i\omega_n)e^{i\omega_n 0^+}\\
     G_{\nu\nu'}(\mathbf{k},i\omega_n) &= \Big[(i\omega_n-\varepsilon_{\mathbf{k}\nu}+\mu)\delta_{\nu\nu'}
      - \Sigma_{\nu\nu'}(\mathbf{k},i\omega_n)\Big]^{-1},\label{Gnu}
  \end{align}
  where $\Sigma_{\nu\nu'}(\mathbf{k},i\omega_n)$ is given by Eq.~\ref{sigma}.
  This non-diagonal density matrix is used instead of the Kohn-Sham
  density matrix to calculate the charge density in \textit{x lapw2 -qdmft}, 
  see Fig.~\ref{fig:sccycle}. Line 4 in Lst.~\ref{densitycorrection} saves
  properties to the hdf5 archive such that they are available in the next
  iteration.
\end{itemize}

Compared to one-shot calculations, the fully charge self-consistent calculations
are in general not much more demanding. As a rough guide, the additional
computational effort due to an increased number of self-consistency
loops is normally around 
50\%. Of course, this extra effort depends largely on the problem at hand
and whether the DMFT corrections to the charge density are relevant or not.

This charge self-consistent implementation also allows the calculation of total
energies from DFT+DMFT. The detailed formulas and their derivation can be found
in Ref.~\cite{triqs_wien2k_full_charge_SC} and references therein. 

For more options of the \py{TRIQS/DFTTools} modules, as well as a complete example
script and tutorial for a charge self-consistent calculation, including total
energy calculation, we refer the reader to the online documentation of
\py{TRIQS/DFTTools}. 

\section{Post-processing}
\label{sec:postprocessing}

\begin{figure}[t]
  \centering
  \includegraphics[width=0.6\textwidth]{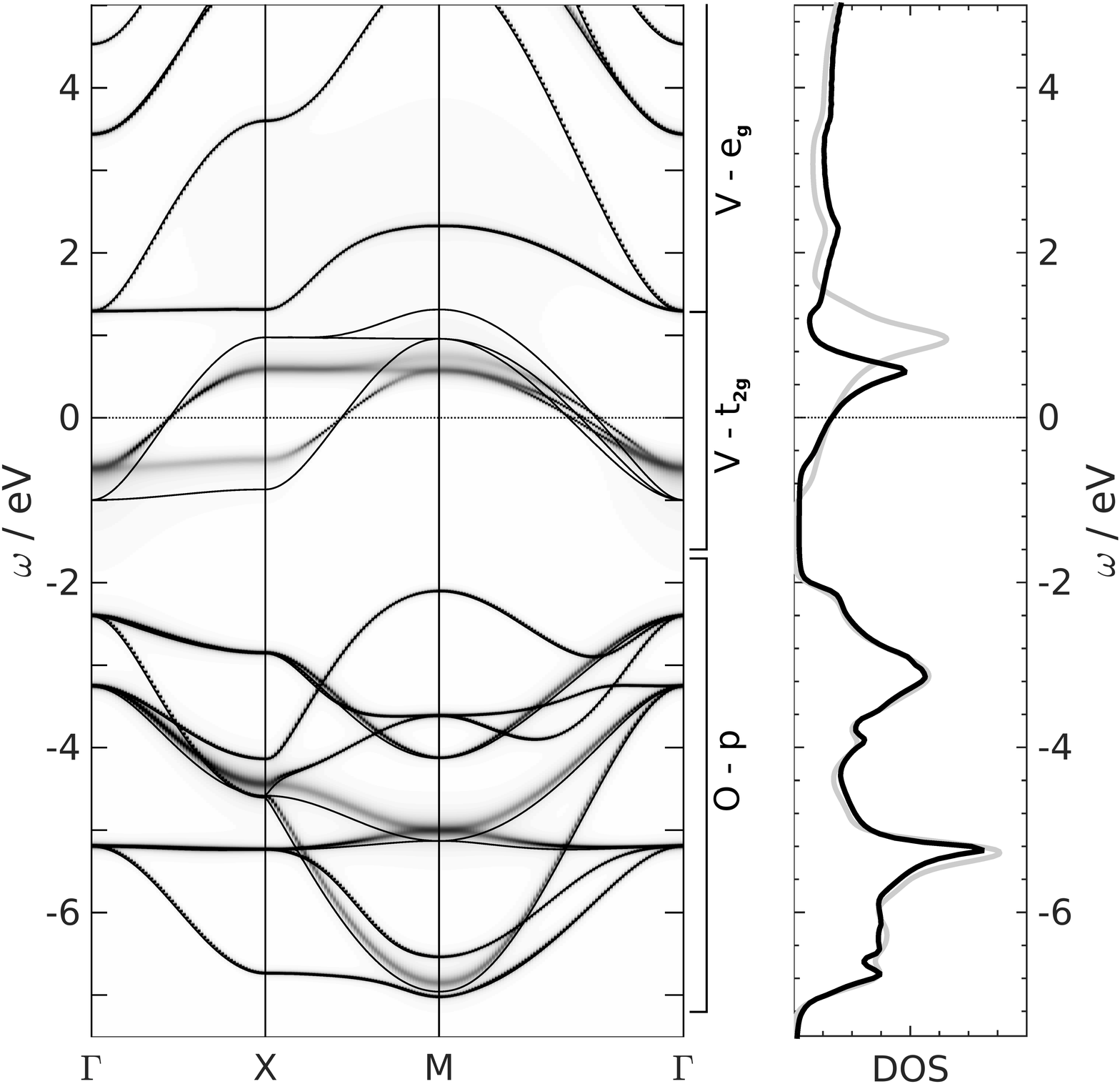}
  \caption{ \label{fig:bands}
          Left: Correlated band structure of SrVO$_3$ (grey-shaded plot) compared
  to the DFT band structure (black lines). Right: Correlated local spectral function (black line) and DFT DOS (grey line).}
\end{figure}

The main output of a DMFT calculation, be it one-shot or fully charge
self-consistent, is the interacting Green's function and the self-energy.
However, in many cases one is interested in other quantities that are more
closely connected to observables of experiments. For that reason \py{TRIQS/DFTTools}
provides methods to further process the data and calculate some physical
properties
\begin{itemize}
  \item (orbitally-projected) density of states
  \item correlated band structures (spectral function $A(k,\omega)$)
  \item transport properties (resistivity, thermopower, optical
    conductivity)
\end{itemize}

For all these methods one needs to use a real-frequency self-energy. In case the
solver provides only quantities on the Matsubara axis, one has to
analytically continue the data, by methods like Pade approximants~\cite{triqs_2015},
the maximum entropy method \cite{jarrell_bayesian_1996}, or others.

\textit{DOS and band structure.} These two quantities are very similar in terms
of equations, since they are both imaginary parts of retarded Green's functions.
For the band structure, we use
\begin{equation}
  A(\mathbf{k},\omega) =
  -\frac{1}{\pi}\mathrm{Tr}\,\mathrm{Im}\,G_{\nu\nu'}(\mathbf{k},\omega+i0^+),
\end{equation}
where the Green's function is defined equivalently to Eq.~\ref{Gnu}, but using a
real-frequency self-energy. The trace is taken in matrix space of Bloch indices $\nu,\nu'$. Summing
this quantity over the first BZ gives the local spectral function.

As an example, we show in Fig.~\ref{fig:bands} a comparison of the DFT and the
DFT+DMFT band structures (left) as well as the local spectral function
and DOS (right) of SrVO$_3$. The
projective Wannier functions were calculated in a large energy window of
$[-8.5,7.5]$\,eV. The impurity problem was solved using the
\py{TRIQS/CTHYB} solver with interaction parameters $U=6.0$\,eV and
$J=0.65$\,eV, and an analytic continuation was performed using stochastic
maximum entropy~\cite{beach_ME}.
 
The thin black lines are the DFT results. One can clearly identify the mass
renormalisation of the t$_{2g}$ bands, which is a bit smaller than 2 in our
calculation. The oxygen $p$, as well as the vanadium e$_g$ states are only
marginally affected. However, due to the hybridisation between vanadium t$_{2g}$
and oxygen $p$ states, the bands between $-7$ and $-2$\,eV acquire a
finite width.

\begin{figure}[t]
  \centering
  \includegraphics[width=0.6\textwidth]{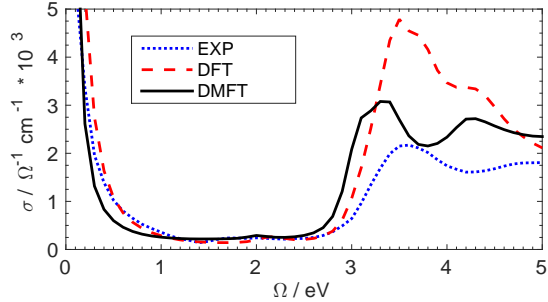}
  \caption{ \label{fig:optcond}
		  Optical conductivity of SrVO$_3$ calculated from the DFT+DMFT result
          (black solid line) and directly from DFT (red dashed line) compared to
          experimental data from Ref.~\cite{PhysRevB.58.4384} (blue dotted line).}
\end{figure}

\textit{Transport properties.} These are evaluated in the Kubo linear-response
formalism~\cite{RevModPhys.78.865} neglecting vertex
corrections~\cite{PhysRevLett.64.1990,PhysRevLett.115.107003}, and at the moment
their computation is based on the Wien2k DFT code. In addition to the
self-energy, one also needs the velocities $v^\alpha(\mathbf k)$ which
are proportional to the
matrix elements of the momentum operator in direction $\alpha=\{x,y,z\}$,
calculated with the Wien2k optics package~\cite{wien2k_optic}. The conductivity
and the Seebeck coefficient for the combination of directions $\alpha\alpha'$ are
defined as  
\begin{equation}
        \label{eq:con_seebeck}
\sigma^{\alpha\alpha'} = \beta e^{2} K_0^{\alpha\alpha'}  \ \ \
\text{and} \ \ \  S^{\alpha\alpha'} =
-\frac{k_B}{|e|}\frac{K_1^{\alpha\alpha'}}{K_0^{\alpha\alpha'}} 
\end{equation}
with $\beta$ the inverse temperature. The kinetic coefficients $K_n^{\alpha\alpha'}$ are given by
\begin{equation}
   K_n^{\alpha\alpha'} = N_{sp} \pi \hbar \int{d\omega
     \left(\beta\omega\right)^n
     f\left(\omega\right)f\left(-\omega\right)\Gamma^{\alpha\alpha'}\left(\omega,\omega\right)}. 
\end{equation}
Here $N_{sp}$ is the spin factor and $f(\omega)$ is the Fermi function. The
transport distribution $\Gamma^{\alpha\alpha'}\left(\omega_1,\omega_2\right)$ is
defined as
\begin{equation}
   \Gamma^{\alpha\alpha'}\left(\omega_1,\omega_2\right) = \frac{1}{V} \sum_k
   \mathrm{Tr}\,\left(v^\alpha(\mathbf{k})A(\mathbf{k},\omega_1)v^{\alpha'}(\mathbf{k})A(\mathbf{k},\omega_2)\right), 
\end{equation}
where $V$ is the unit cell volume. In multi-band systems the velocities
$v^\alpha(\mathbf{k})$ and the spectral function $A(\mathbf{k},\omega)$ are
Hermitian matrices in the band indices $\nu$ and $\nu'$.  The frequency-dependent optical
conductivity is given by
\begin{equation}
    \sigma^{\alpha\alpha'}(\Omega) = N_{sp} \pi e^2 \hbar \int{d\omega
    \Gamma^{\alpha\alpha'}(\omega+\Omega,\omega)
    \frac{f(\omega)-f(\omega+\Omega)}{\Omega}}. 
\end{equation}

A related implementation, also using full dipole matrix elements but
in the framework of maximally localised Wannier
functions, has recently been published as the woptic package~\cite{Assmann2015}.

We illustrate the calculation of transport properties again for the example of
SrVO$_3$. We apply the same setup as for the calculation of the band structure
given above. For the $\mathbf{k}$-summation, we use 4495 $\mathbf{k}$-points in
the irreducible BZ. At $T=290$\,K we obtain a DC resistivity of
$47$\,$\mu\Omega$cm, which compares 
reasonably well with the experimental value of $70$\,$\mu\Omega$cm~\cite{inoue}.
The Seebeck coefficient, $-8$\,$\mu$V/K in
our calculation, agrees remarkably well with the experimental room temperature
value of about $-11$\,$\mu$V/K~\cite{JCCS.14.247}. The same holds for the
optical conductivity (Fig.~\ref{fig:optcond}) obtained from DFT+DMFT (black solid
line) when compared to experimental results (blue dotted line).  To show the
effect of the correlations, we also present results without 
self-energy (red dashed line), which corresponds to evaluating the optical
conductivity directly from DFT.  Sometimes, scattering processes not included in
plain DFT+DMFT calculations (e.g., phonons, impurities, or non-local fluctuations)
are important. In this case, the calculated resistivity constitutes a lower
bound to the expected experimental value. Note that for all transport
calculations the convergence with the number of $\mathbf{k}$-points is
essential and has to be checked carefully.

Further details on the use of these tools, including all options and parameters,
are given in the online documentation. 

\section{Getting started}
\label{sec:starting}

Detailed information on the installation procedure can be found on the
\py{TRIQS/DFTTools} website and current issues and updates are available on
GitHub.

\subsection{Obtaining \py{TRIQS/DFTTools}}

The \py{TRIQS/DFTTools} source code is available publicly and can be obtained by
cloning the repository on the GitHub website at
\href{https://github.com/TRIQS/dft\_tools}{https://github.com/TRIQS/dft\_tools}.
As the TRIQS project, including its applications, is continuously improving, we
strongly recommend users to obtain TRIQS and its applications, including
\py{TRIQS/DFTTools}, from GitHub. Bugfixes to possible issues are also applied to the
GitHub source.

For users that wish to have a more stable version of the code,
without the latest bugfixes and functionalities,
we suggest downloading the latest tagged version from the GitHub releases page
\href{https://github.com/TRIQS/dft\_tools/releases}{https://github.com/TRIQS/dft\_tools/releases}. 

\subsection{Installation}

Once the TRIQS library has been installed properly, the installation of
\py{TRIQS/DFTTools} is straightforward. As for TRIQS we use the \verb#cmake# tool to
configure, build and test the application. With the TRIQS library installed in
\verb#/path/to/TRIQS/install/dir#, the \py{TRIQS/DFTTools} application can be
installed by
\begin{verbatim}
$ git clone https://github.com/TRIQS/dft_tools.git src
$ mkdir build_dft_tools && cd build_dft_tools
$ cmake -DTRIQS_PATH=/path/to/TRIQS/install/dir ../src
$ make
$ make test
$ make install
\end{verbatim}
This will first download the code to the \py{src} directory, then build, test,
and install the application in the same location as the TRIQS library.

\subsection{Citation policy}

We kindly request that the present paper is cited in any published work using
the \py{TRIQS/DFTTools} package. Furthermore, we request that the original
works to this package, Refs.~\cite{triqs_wien2k_interface} for the
first implementation and
\cite{triqs_wien2k_full_charge_SC} for full charge self-consistency,
are cited accordingly, too. In addition, 
since this package is based on the TRIQS library, a citation
to~\cite{triqs_2015} is also requested, along with the appropriate citation to
any solver used, e.g. \py{TRIQS/CTHYB}~\cite{cthyb_2015}. 

\subsection{Contributing}

\py{TRIQS/DFTTools}, as an application of TRIQS, is an open source project. We
appreciate feedback and contributions from the user community to this package.
Issues and bugs can be reported on the GitHub website
(\href{https://github.com/TRIQS/dft\_tools/issues}{https://github.com/TRIQS/dft\_tools/issues}).
Before starting major contributions, please coordinate with the team of
\py{TRIQS/DFTTools} developers.

\section{Summary}
\label{sec:summary}

In summary we presented the \py{TRIQS/DFTTools} package, an application based on the
TRIQS library. This package provides the necessary tools for setting up DFT+DMFT
calculations. We have shown simple examples demonstrating how the code can be
used to calculate Wannier functions, spectral functions, as well as transport
properties of the prototypical material SrVO$_3$. 

\section{Acknowledgements}
\label{sec:acknowledgements}

We gratefully acknowledge discussions, comments, and feedback from the
user community, and ongoing collaborations with A.~Georges and S.~Biermann. 
M.~Aichhorn, M.~Zingl, and G.~J.~Kraberger acknowledge financial support from
the Austrian Science Fund (Y746, P26220, F04103), and great hospitality at
Coll\`ege de France, \'Ecole Polytechnique, and CEA Saclay. 
L.~Pourovskii
acknowledges the financial support of the Ministry of Education and Science of
the Russian Federation in the framework of Increase Competitiveness Program of
NUST MISiS (No. K3-2015-038). 
L.~Pourovskii and V.~Vildosola acknowledge
financial support from ECOS-A13E04. 
O.~E.~Peil acknowledges support from the Swiss National Science
Foundation (program NCCR-MARVEL). 
P.~Seth acknowledges support from
ERC Grant No. 617196-CorrelMat. 
O.~Parcollet and P.~Seth
acknowledge support from ERC Grant No. 278472-MottMetals.
J.~Mravlje acknowledges discussions with J.~Tomczak and support of
Slovenian Research Agency (ARRS) under Program P1-0044.  

\bibliographystyle{elsarticle-num}
\bibliography{refs.bib}

\end{document}